\documentclass[journal]{IEEEtran}
\IEEEoverridecommandlockouts

\usepackage{cite}
\usepackage{amsmath,amssymb,amsfonts}
\usepackage{mathtools}
\usepackage{amsthm}
\usepackage{algorithm}
\usepackage{algpseudocode}
\usepackage{algorithm}
\usepackage{algpseudocode} 
\usepackage{etoolbox}
\usepackage{graphicx,color}
\usepackage{textcomp}
\usepackage[dvipsnames,rgb]{xcolor}
\usepackage{makecell}
\usepackage{float}
\usepackage{tikz}
\usetikzlibrary{arrows.meta, positioning, shapes.geometric,decorations.pathreplacing}
\usetikzlibrary{patterns}
\usepackage{siunitx}
\usepackage{tabularx}
\usepackage{colortbl}
\usepackage{multirow}
\usepackage{array}
\usepackage{acronym}

\usepackage{subcaption}
\usepackage{booktabs}
\usepackage{hyperref} 
\usepackage[capitalize,noabbrev]{cleveref}
\usepackage[textsize=tiny]{todonotes}
\usepackage{latexsym}
\usepackage{gensymb}
\usepackage{enumitem}
\usepackage{esvect}
\usepackage{comment}
\usepackage[normalem]{ulem}
\usepackage{soul}

\input{defs/macros}
\theoremstyle{plain}
\newtheorem{theorem}{Theorem}

\theoremstyle{definition}

\theoremstyle{remark}
\newtheorem{remark}[theorem]{Remark}

\definecolor{nyuviolet}{RGB}{87, 6, 140}

\def\BibTeX{{\rm B\kern-.05em{\sc i\kern-.025em b}\kern-.08em
    T\kern-.1667em\lower.7ex\hbox{E}\kern-.125emX}}

\newacro{ACDD}{Alamouti with cyclic delay diversity}
\newacro{URLLC}{ultra-reliable low-latency communications}
\newacro{3GPP}{third generation partnership project}
\newacro{PHY}{physical layer}
\newacro{MIMO}{multiple-input multiple-output}
\newacro{MU-MIMO}{multi-user multiple-input multiple-output}
\newacro{SIMO}{single-input multiple-output}
\newacro{MISO}{multiple-input single-output}
\newacro{SISO}{single-input single-output}
\newacro{MRC}{maximum-ratio combining}
\newacro{SNR}{signal-to-noise ratio}
\newacro{CP}{cyclic prefix}
\newacro{CDD}{cyclic delay diversity}
\newacro{FSC}{frequency-selective channel}
\newacro{STC}{space-time coding}
\newacro{FFT}{fast Fourier transform}
\newacro{LMMSE}{linear minimum mean-squared error}
\newacro{CFAC}{cross frequency AoA consistency}
\newacro{FER}{frame error rate}
\newacro{OFDM}{orthogonal frequency division multiplexing}
\newacro{OCDM}{orthogonal chirp division multiplexing}
\newacro{RMS}{root mean square}
\newacro{DS}{delay spread}
\newacro{FSC}{frequency-selective channel}
\newacro{CSI}{channel state information}
\newacro{LMMSE-PIC}{linear minimum mean squared error with parallel interference cancellation}
\newacro{PFE}{perfect-feedback equalizer}
\newacro{FD}{full-duplex}
\newacro{PDP}{power delay profile}
\newacro{PDF}{probability density function}
\newacro{DFT}{discrete Fourier transform}
\newacro{SDFT}{sparse DFT}
\newacro{ICI}{inter-carrier interference}
\newacro{OTFS}{orthogonal time frequency space}
\newacro{AWGN}{additive white Gaussian noise}
\newacro{SWH}{sparse Walsh-Hadamard}
\newacro{LLR}{log-likelihood ratio}
\newacro{PMF}{probability mass function}
\newacro{CRC}{cyclic redundancy check}
\newacro{PAM}{pulse amplitude modulation}
\newacro{QAM}{quadrature amplitude modulation}
\newacro{FWHT}{fast Walsh-Hadamard transform}
\newacro{MAP}{maximum a-posteriori}
\newacro{SC}{specular component}
\newacro{CFO}{carrier frequency offset}
\newacro{ISI}{inter-symbol interference}
\newacro{ZP}{zero-padding}
\newacro{EVD}{eigenvalue decomposition}
\newacro{BCJR}{Bahl, Cocke, Jelinek, and Raviv}
\newacro{WHT}{Walsh-Hadamard transform}
\newacro{APP}{a-posteriori probability}
\newacro{SILE-EPIC}{self-iterated linear equalizer with expectation propagation}
\newacro{EP}{expectation propagation}
\newacro{i.i.d.}{independent and identically distributed}
\newacro{CWCU}{component wise conditionally unbiased}
\newacro{MSE}{mean squared error}
\newacro{EXIT}{extrinsic information transfer}
\newacro{MI}{mutual information}
\newacro{PAPR}{peak-to-average power ratio}
\newacro{DFT-s}{discrete Fourier transform-spread}
\newacro{AMP}{approximate message passing}
\newacro{GAMP}{generalized \ac{AMP}}
\newacro{VAMP}{vector \ac{AMP}}
\newacro{RSC}{recursive systematic convolutional}
\newacro{QPSK}{quadrature phase-shift keying}
\newacro{CFAR}{constant false alarm rate}
\newacro{PD}{probability of detection}
\newacro{PFA}{probability of false alarm}
\newacro{RV}{random variable}
\newacro{CDF}{cumulative distribution function}
\newacro{HD-ZP}{half-duplex ZP}
\newacro{FD-CP}{full-duplex ZP}
\newacro{DFRC}{dual-function radar communication}
\newacro{SINR}{signal-to-interference noise ratio}
\newacro{ISAC}{integrated sensing and communication}
\newacro{SI}{self-interference}
\newacro{RSI}{residual self-interference}
\newacro{ADC}{analog-to-digital converter}
\newacro{DAC}{digital-to-analog converter}
\newacro{ED}{energy-detection}
\newacro{IDFT}{inverse discrete Fourier Transform}
\newacro{SFFT}{symplectic finite Fourier transform }
\newacro{CRB}{Cram{\'{e}}r-Rao bound}
\newacro{ZC}{Zadoff-Chu}
\newacro{RMSE}{root mean square error}
\newacro{MMSE}{minimum mean-square error}
\newacro{UW}{unique word}
\newacro{GFDM}{generalized frequency division multiplexing}
\newacro{RRC}{root-raised cosine}
\newacro{UB}{upper bound}
\newacro{CEF}{channel estimation field}
\newacro{TRX}{transceiver}
\newacro{IF}{intermediate frequency}
\newacro{RF}{radio frequency}
\newacro{FPGA}{field programmable gate arrays}
\newacro{SDR}{software-defined radio}
\newacro{UWB}{ultra wideband}
\newacro{FR3}{frequency range 3}
\newacro{PCB}{printed circuit board}
\newacro{SMA}{SubMiniature version A}
\newacro{MUSIC}{multiple signal classification}
\newacro{CIR}{channel impulse response}
\newacro{FR}{Frequency Range}
\newacro{mmWave}{millimeter wave}
\newacro{LoS}{line-of-sight}
\newacro{AoD}{angle-of-departure}
\newacro{ESNR}{estimation SNR}
\newacro{AoA}{angle-of-arrival}
\newacro{SDNR}{signal-to-DMC-noise ratio}
\newacro{ULA}{uniform linear array}
\newacro{DMC}{dense multipath component}
\newacro{ML}{maximum-likelihood}
\newacro{IFFT}{inverse fast Fourier transform}
\newacro{LM}{Levenberg-Marquardt}
\newacro{ACF}{autocorrelation function}
\newacro{UWB}{ultra-wideband}
\newacro{SLAM}{simultaneous localization and mapping}
\newacro{STO}{sampling time offset}
\newacro{GLRT}{generalized likelihood ratio test}
\newacro{FD-ED}{frequency-domain energy detectors}
\newacro{IOU}{intersection over union}
\newacro{FLOP}{floating point operation}
\newacro{FLOPS}{floating point operations per second}
\newacro{LTBF}{long-term beamforming}
\newacro{UE}{user equipment}
\newacro{SRS}{sounding reference signal}
\newacro{BW}{bandwidth}
\newacro{RB}{resource block}
\newacro{RE}{resource element}
\newacro{BS}{base station}
\newacro{CG}{conjugate-gradient}
\newacro{SRAM}{static random access memory}
\newacro{DSP}{digital signal processor}
\newacro{VLSI}{very large-scale integration}
\newacro{NR}{New Radio}
\newacro{ASIC}{application-specific integrated circuit}
\newacro{gNB}{next-generation NodeB}
\newacro{SCS}{subcarrier spacing}
\newacro{WMMSE}{weighted minimum mean square error}
\newacro{WSR}{weighted sum-rate}
\newacro{ZF}{zero-forcing}
\newacro{UMB}{upper mid-band}
\newacro{DM-RS}{Demodulation Reference Signal}
\newacro{RC-EVD}{randomized complex eigenvalue decomposition}
\newacro{GEMM}{generalized matrix multiplication}
\newacro{SVD}{singular value decomposition}

\begin{document}

\setlength{\abovedisplayskip}{3pt plus 1pt minus 1pt}
\setlength{\belowdisplayskip}{3pt plus 1pt minus 1pt}
\setlength{\abovedisplayshortskip}{1pt plus 1pt}
\setlength{\belowdisplayshortskip}{1pt plus 1pt minus 1pt}
\setlength{\textfloatsep}{5pt plus 1pt minus 2pt}
\setlength{\floatsep}{3pt plus 1pt minus 1pt}
\setlength{\intextsep}{3pt plus 1pt minus 1pt}
\setlength{\dbltextfloatsep}{5pt plus 1pt minus 2pt}
\setlength{\dblfloatsep}{3pt plus 1pt minus 1pt}
\title{Low-rank Preconditioning in Beamspace Domain For Massive MU-MIMO Long-Term Beamforming 
}

\author{
Amirreza Kiani$^{\diamond}$,
Ali Rasteh$^*$,
Marco Mezzavilla$^{\diamond}$,
and Sundeep Rangan$^*$ \\
    $^\diamond$Dipartimento di Elettronica, Informazione e Bioingegneria (DEIB), Politecnico di Milano, Milan, Italy\\
	$^*$NYU WIRELESS, NYU Tandon School of Engineering, New York, USA \\
	Email: amirreza.kiani@polimi.it, ar7655@nyu.edu, marco.mezzavilla@polimi.it, srangan@nyu.edu 
}

\maketitle

\thispagestyle{empty}
\pagestyle{empty}

\acresetall
\begin{abstract}
Long-term beamforming substantially reduces the channel
estimation and inversion overhead of conventional massive MU-MIMO
receivers; yet, its construction still hinges on the inversion of a
large Hermitian matrix, whose condition number deteriorates with the
per-user SNR dynamic range. When this inversion is approximated in
hardware via the conjugate gradient (CG) algorithm, the deterioration
directly inflates the iteration count and, consequently, the energy
and latency budget. We propose a hardware-friendly low-rank
preconditioning framework that targets exactly this bottleneck. The
preconditioner is constructed from the top eigenpairs of the
long-term covariance matrix through a randomized complex eigenvalue
decomposition (RC-EVD), whose inner QR factorizations are realized via
a Cholesky-based scheme (QRC), confining the dominant cost to generalized
matrix multiplication (GEMM) and
small triangular solves that map naturally onto systolic arrays. We
further show that performing the preconditioned CG inversion in the
beamspace domain induces sparsification of the system matrix and
provides additional convergence acceleration at negligible
transformation cost. Ray-tracing simulations confirm that the joint
scheme reduces the required CG iteration count by two to three while
matching the post-equalization SINR of the exact inversion.
\end{abstract}


 \section{Introduction}
Massive \ac{MIMO} is a cornerstone of the capacity gains delivered by 5G
\cite{larsson2014massive,jin2023massive}, and ongoing efforts target a
substantial increase in the number of antenna elements at the base
station \cite{nokia2025massiveMIMO}. Larger arrays not only improve
spectral efficiency but also strengthen interference
suppression \cite{jia2025joint} and unlock operation across wider
bandwidths \cite{akrout2023bandwidth}. These gains, however, are
conditioned on baseband architectures that can keep up with the
resulting dimensionality \cite{dai2021scalable}.

\ac{LTBF} \cite{lozano2007long} is a particularly attractive route in
this direction, as it exploits the slowly varying second-order
statistics of the channel to perform spatial interference suppression
at a coherence-block granularity rather than per resource element.
Both the channel-estimation overhead and the inversion workload are
thus amortized over many subcarriers and slots, which is essential for
a scalable hardware realization.

Even after this amortization, the construction of the long-term
projection in a cellular \ac{MU-MIMO} uplink still requires the inversion
of an $N \times N$ Hermitian matrix
\cite{rasteh2025scalable}, where $N$ denotes the number of
base-station antennas. At the array sizes envisioned for 5G-Advanced
and 6G, a closed-form inversion is no longer viable, and the operation
must be approximated in hardware. Iterative methods, and \ac{CG} in particular, are the standard choice in this regime
because of their regular data flow and low control overhead
\cite{yin2015vlsi,dai2021scalable}. The number of \ac{CG} iterations,
however, scales with the square root of the condition number of the
    target matrix, which itself grows with the per-user \ac{SNR} dynamic
range, a quantity that uplink power control can only partially
constrain \cite{rasteh2025scalable}. Reducing this iteration count
without compromising post-equalization performance is therefore the
key lever for a hardware-efficient \ac{LTBF} implementation, and the
focus of this work.

\textbf{Contributions:} We propose a hardware-friendly preconditioning
framework that targets exactly this lever. The contributions are
threefold:
\begin{enumerate}
    \item We introduce a low-rank preconditioner motivated by the
    eigenvalue clustering of the long-term covariance matrix and
    apply it to \ac{CG} via the Sherman–Morrison–Woodbury identity at
    $\mathcal{O}(qN^2)$ cost per iteration rather than
    $\mathcal{O}(N^3)$.
    \item We construct the preconditioner through a truncated
    \ac{RC-EVD}, in which the
    inner tall-and-skinny QR factorizations are replaced by a
    Cholesky-based variant (QRC). The resulting kernel is dominated by
    \ac{GEMM} and small triangular solves, both of which map naturally onto
    systolic-array architectures.
    \item We show that performing the preconditioned \ac{CG} inversion in
    the beamspace domain induces sparsification of the system matrix
    and further accelerates convergence at negligible transformation
    cost.
\end{enumerate}
The framework is validated on ray-traced channels, where it recovers
near-exact post-beamforming \ac{SINR} with substantially fewer \ac{CG}
iterations than the unpreconditioned baseline.

 \section{System Model}
\label{sec:sys-model}

\subsection{Multi-User MIMO Uplink Signal Model}
We consider a multi-user \ac{MIMO} uplink in which $N_{\subsf{UE}}$
\acp{UE} share a common time--frequency resource, and each \ac{UE}
transmits $N_s$ data streams. Adopting the standard
\ac{OFDM} formulation~\cite{heath2018foundations}, the received signal
at the base station on subcarrier $n$ of \ac{OFDM} symbol $k$ is
\begin{equation} \label{eq:mumimo}
    \bs{y}[n,k] \;=\; \sum_{i=1}^{N_{\subsf{UE}}}
        \bs{H}_i[n,k]\, \bs{x}_i[n,k] \;+\; \bs{w}[n,k],
\end{equation}
where $\bs{y}[n,k] \in \mathbb{C}^{N}$ is the received vector across
the $N$ base-station antennas, $\bs{H}_i[n,k] \in
\mathbb{C}^{N \times N_s}$ is the effective channel matrix from
\ac{UE} $i$ (incorporating any user-side precoder), $\bs{x}_i[n,k]$
collects its $N_s$ transmitted symbols, and $\bs{w}[n,k]$ is the
additive receiver noise. Letting $\mc{E}_{x_i}$ denote the
per-symbol transmit energy of \ac{UE} $i$, we assume i.i.d. symbols
across streams,
\begin{equation} \label{eq:xvar}
    \Exp\!\left[\, \bs{x}_i[n,k]\,\bs{x}_i\herm[n,k]\,\right]
    \;=\; \frac{\mc{E}_{x_i}}{N_s}\, \bs{I}.
\end{equation}

\subsection{Multi-User Long-Term Beamforming}
\label{sec:ltbf}
The principle behind multi-user \ac{LTBF} is to suppress inter-user
interference through a projection that depends only on the
long-term spatial statistics of the channel. For each \ac{UE} $i$,
the receiver applies
\begin{equation} \label{eq:zproj}
    \bs{z}_i[n,k] \;=\; \bs{G}_i\, \bs{y}[n,k],
\end{equation}
where $\bs{G}_i \in \mathbb{C}^{r \times N}$, with $r < N$, projects
the received signal onto an $r$-dimensional subspace in which the
contributions of the interfering users are nominally
attenuated. Because $\bs{G}_i$ is computed from second-order
statistics rather than instantaneous channel realizations, it is
held constant across all subcarriers $n$ and over the entire
long-term coherence interval. As a result, the matrix inversion of
size $N \times N$ required for the construction of
$\bs{G}_i$~\cite{rasteh2025scalable} is performed once per coherence
window and amortized across the full bandwidth, in stark contrast to
instantaneous beamforming, where an inversion is carried out per
\ac{UE} and per resource element.
The projection in \eqref{eq:zproj} captures only the spatial
structure of the interference and does not equalize the
frequency-selective small-scale fading. Frequency-selective
equalization is therefore performed after the projection, on a
per-subcarrier basis, by a conventional \ac{MMSE} or \ac{ZF}
front end.

\section{Conjugate Gradient for Matrix Inverse Approximation}
\label{sec:cg}

The construction of the long-term projector~$\bs{G}_i$
in~\eqref{eq:zproj} reduces to the inversion of a Hermitian
positive-definite matrix
\begin{equation} \label{eq:Q-def}
    \bs{Q} \;=\; \bs{I} + \sum_{i=1}^{N_{\subsf{UE}}} \alpha_i\,
    \bar{\bs{R}}_i,
    \qquad
    \alpha_i \;\triangleq\; \frac{\mc{E}_{x_i}}{N_0 N_s},
\end{equation}
where $\bar{\bs{R}}_i$ is the long-term spatial correlation of
$\bs{H}_i$, and $N_0$ is the noise power spectral
density~\cite{rasteh2025scalable}. Since closed-form inversion of
$\bs{Q}$ is intractable at the array sizes targeted by 5G-Advanced
and 6G, we approximate $\bs{Q}^{-1}$ iteratively. Among the
candidates, the \ac{CG} algorithm is particularly well suited to hardware implementation because its dominant kernels are matrix–vector and matrix–matrix multiplications, both of which map
naturally onto regular dataflow architectures such as systolic
arrays~\cite{albreem2021low,fang2025finite,rasteh2025spatial}.

The \ac{CG} method seeks $\bs{X} \approx \bs{Q}^{-1}$ by solving
\begin{equation}
    \bs{Q}\,\bs{X} \;=\; \bs{I},
\end{equation}
and the iteration is terminated as soon as the relative residual
falls below a target tolerance,
\begin{equation} \label{eq:apperror}
    \|\, \bs{Q}\,\bs{X} - \bs{I}\,\| \;<\; \epsilon.
\end{equation}
For a Hermitian positive-definite system, the number of iterations
required to reach an accuracy $\epsilon$ is bounded
by~\cite{saad2003iterative}
\begin{equation} \label{eq:convergCG}
    k \;=\; \mathcal{O}\!\left(\sqrt{\kappa(\bs{Q})}\,
        \log\!\tfrac{1}{\epsilon}\right),
\end{equation}
where $\kappa(\bs{Q}) = \lambda_{\max}(\bs{Q})/\lambda_{\min}(\bs{Q})$
is the spectral condition number of $\bs{Q}$.

\theoremstyle{remark}
\begin{remark}\label{rem:cond-snr}
From~\eqref{eq:Q-def}, the smallest eigenvalue of $\bs{Q}$ is lower
bounded by unity due to the identity term, while its largest
eigenvalue grows linearly with the effective \acp{SNR} $\{\alpha_i\}$
through the rank-aggregated term $\sum_i \alpha_i \bar{\bs{R}}_i$.
Consequently, $\kappa(\bs{Q})$ scales linearly with the per-user
\ac{SNR} dynamic range, and high-\ac{SNR} regimes yield an
increasingly ill-conditioned system that, by~\eqref{eq:convergCG},
inflates the \ac{CG} iteration count.
\end{remark}

The accuracy criterion in~\eqref{eq:apperror} translates directly
into a guarantee on the post-equalization \ac{SINR}. Let $\gamma_i^0$
denote the per-user \ac{SINR} achieved by the exact \ac{LTBF}
projection (i.e.\ when $\bs{Q}^{-1}$ is computed without
approximation), and let $\gamma_i$ denote the \ac{SINR} obtained
when $\bs{X}$ is used in place of $\bs{Q}^{-1}$. As shown
in~\cite{rasteh2025scalable}, the two are related, on any resource
element $(n,k)$, by
\begin{equation} \label{eq:sinr_error}
    \gamma_i
    \;\geq\;
    \frac{\gamma_i^0\,(1-\epsilon)^2}
         {(1+\epsilon)^2 + 4\epsilon\, \Exp(\gamma_i^0)},
\end{equation}
where the expectation is taken over the small-scale fading
realizations $c_i[n,k]$. Note that $\gamma_i^0$ is itself a function
of the per-user \acp{SNR} $\{\alpha_i\}$ defined in~\eqref{eq:Q-def},
so~\eqref{eq:sinr_error} establishes a direct link between the
inversion accuracy $\epsilon$, the operating \ac{SNR}, and the
loss incurred by the approximate filter — thereby motivating the
preconditioning strategy developed in the next section.

\section{Low-rank Preconditioning in the Beamspace Domain}

Preconditioning is one of the most well-known strategies adapted to effectively reduce the number of required \ac{CG} iterations, thereby alleviating the associated computational burden~\cite{fang2025finite}, \cite{liu2020energy}. In this paper, we aim to propose an effective and hardware-friendly preconditioning technique to alleviate the negative impact of high \ac{SNR} dynamic range in our inversion approximation.

Preconditioning typically consists of two main components: (i) the construction of the preconditioner based on target problem properties and (ii) its application during the \ac{CG} iterations.

Several approaches exist for constructing preconditioners $\bs{M}$. In this work, we adopt a low-rank preconditioning strategy motivated by the eigenvalue structure of the target matrix $\bs{Q}$. We begin by examining the structure of the target matrix and outlining the construction of the preconditioner, followed by its application in the \ac{CG} based approximation.

\subsection{Preconditioner Construction and Application in CG}
Ideally, a good preconditioner satisfies $\bs{M} \approx \bs{Q}^{-1}$, which improves the conditioning of the system and accelerates convergence~\cite{diouane2026spectral}. 
We start by investigating the eigenvalue clustering of matrix $\bf{Q}$. We show that this matrix can be viewed as a low-rank perturbation of the identity, leading to eigenvalue clustering. Hence, we build our preconditioner motivated by this property. 

\theoremstyle{remark}
\begin{remark}
The result of multiplication $\bf{H}\bf{H}^H$ is a low-rank matrix since its non-zero eigenvalues are identical to those of $\bf{H}^H \bf{H}$. As a consequence, $N - \sum_i N_s$ eigenvalues of $\bf{Q}$ are clustered around 1, while the remaining eigenvalues are relatively larger.
\end{remark}

Now let
\begin{equation}
\hat{\bs{Q}} = \sigma^2 \bs{I} + \bs{U}_q (\bs{\Lambda}_q - \sigma^2 \bs{I}_q) \bs{U}_q^H
\end{equation}
be an approximation of $\bs{Q}$, where
\begin{itemize}
\item $\bs{U}_q \in \mathbb{C}^{N \times q}$ contains the $q$ leading eigenvectors,
\item $\bs{\Lambda}_q = \mathrm{diag}(\lambda_1, \dots, \lambda_q)$, where $\lambda_i$ are the eigenvalues associated with the leading eigenvectors, ordered in descending order,
\item $\sigma^2$ is a scalar representing the average spectral scale.
\end{itemize}
The matrices $\bs{U}_q$ and $\bs{\Lambda}_q$ can be obtained using truncated \ac{EVD} methods. 
The scalar $\sigma^2$ can be computed as $\sigma^2 = \operatorname{Re}\!\left( \frac{\operatorname{tr}(\bs{Q})}{N} \right)$.

Using the Sherman–Morrison–Woodbury identity, we obtain:
\begin{equation} \label{eq:lowrankConstrucion}
\bs{M} = \hat{\bs{Q}}^{-1}  = \sigma^{-2} \bs{I}
- \bs{U}_q (\bs{\Lambda}_q - \sigma^2 \bs{I}_q)\,(\sigma^2 \bs{\Lambda}_q)^{-1} \bs{U}_q^H.
\end{equation}
Preconditioner $\bs{M}$ can be explicitly employed in a compact matrix form of the \ac{CG} algorithm, as shown in Algorithm~\ref{alg:pmatrix_cg_inverse}, or implicitly using


\begin{equation} \label{eq:impl}
\bs{M} {\bs{R}^{(i+1)}}
= \sigma^{-2}{\bs{R}^{(i+1)}}
- \bs{U}_q \bs{D} \left(\bs{U}_q^H \bs{R}^{(i+1)}\right),
\end{equation}
where $\bs{D} = \sigma^{-2}\bs{I}_q - \bs{\Lambda}_q^{-1}$. Evaluated right-to-left, this is more efficient than the explicit matrix formulation, as it reduces the computational complexity of the preconditioner application from $\mathcal{O}(N^3)$ down to $\mathcal{O}(q N^2)$.

\begin{algorithm}[t]
\footnotesize
\caption{Preconditioned CG for $\bs{Q} \bs{X} = \bs{I}$}
\label{alg:pmatrix_cg_inverse}
\begin{algorithmic}[1]

\Require ${Q} \in \mathbb{C}^{N \times N}$, preconditioner ${M}$, initial guess ${X}^{(0)} \in \mathbb{C}^{N \times N}$
\Ensure Approximate inverse ${X} \approx {Q}^{-1}$

\State ${R}^{(0)} \gets {I} - {Q} {X}^{(0)}$
\State ${Z}^{(0)} \gets {M} {R}^{(0)}$
\State ${P}^{(0)} \gets {Z}^{(0)}$

\For{$i = 0,1,2,\dots k'$}

    \State ${S}^{(i)} \gets {Q} {P}^{(i)}$

    \State ${\alpha}^{(i)} \gets \mathrm{diag}\!\left(
    \frac{(r^{(i)}_j)^H z^{(i)}_j}{(p^{(i)}_j)^H s^{(i)}_j}
    \right)$

    \State ${X}^{(i+1)} \gets {X}^{(i)} + {P}^{(i)} \bs{\alpha}^{(i)}$

    \State ${R}^{(i+1)} \gets {I} - {Q} {X}^{(i+1)}$

    \State ${Z}^{(i+1)} \gets {M} {R}^{(i+1)}$

    \State ${\beta}^{(i)} \gets \mathrm{diag}\!\left(
    \frac{(r^{(i+1)}_j)^H z^{(i+1)}_j}{(r^{(i)}_j)^H z^{(i)}_j}
    \right)$

    \State ${P}^{(i+1)} \gets {Z}^{(i+1)} + {P}^{(i)} {\beta}^{(i)}$

\EndFor

\end{algorithmic}
\end{algorithm}
The computational burden of preconditioner construction is primarily dominated by the truncated \ac{EVD}. Accordingly, in section~\ref{sec:EVD}, we show how this operation can be done in an efficient and hardware friendly way. 

Note that with respect to Eq. \ref{eq:convergCG}, the convergence rate of preconditioned \ac{CG} can be bounded by the condition number $\kappa(\bar{\bs{Q}}) \ll \kappa({\bs{Q}})$.
Here, the matrix $\bar{\bs{Q}}$ is defined as
\begin{equation}
\bar{\bs{Q}} \;=\; \bs{L}\,\bs{Q}\,\bs{L}^H,
\end{equation}
where $\bs{M} = \bs{L}\bs{L}^H$. With this factorization, applying the \ac{CG} method to the preconditioned system
\begin{equation}
\bs{M}^{-1}\bs{Q}\,\bs{x} = \bs{M}^{-1}\bs{b}
\end{equation}
is equivalent to applying \ac{CG} to the symmetrically preconditioned system
\begin{equation}
\bar{\bs{Q}}\,\bs{y} = \bar{\bs{b}}, \qquad 
\bs{y} = \bs{L}^{-H}\bs{x}, \quad 
\bar{\bs{b}} = \bs{L}\bs{b}.
\end{equation}
Since the dominant eigenmodes of $\bs{Q}$ are explicitly captured in the rank-$q$ approximation, the spectrum of the preconditioned system is expected to be more tightly clustered around unity. This clustering effect reduces the effective condition number of the system, which improves \ac{CG} convergence~\cite{diouane2026spectral}.


\subsection{Preconditioning and Inversion in Beamspace Domain}

Instead of performing preconditioning and matrix inversion in the antenna domain, we operate in the beamspace domain~\cite{sayeed2013beamspace}, which allows us to exploit the inherent sparsity of the channel representation. A rigorous analytical characterization of the advantages of beamspace-domain processing is beyond the scope of this work and is left for future investigation. Instead, we provide simulation results to demonstrate the benefits of operating in the beamspace domain. 

We construct a unitary \ac{DFT} matrix $\bs{F} \in \mathbb{C}^{N \times N}$ for a planar array with dimensions ${T \times T}$ where $T = \sqrt{N}$ using a separable structure. Let $\bs{F}_x, \bs{F}_y \in \mathbb{C}^{T \times T}$ be unitary \ac{DFT} matrices defined as
\begin{equation}
    \bs{F}_x = \frac{1}{\sqrt{T}}\mathrm{FFT}(\bs{I}), \quad
    \bs{F}_y = \frac{1}{\sqrt{T}}\mathrm{FFT}(\bs{I}).
\end{equation}
Then, the beamspace transformation matrix is given by
\begin{equation}
    \bs{F} = \bs{F}_x \otimes \bs{F}_y.
\end{equation}
where $\bs{I}$ is the identity matrix and $\otimes$ denotes the Kronecker product. The matrix $\bs{Q}$ is then transformed into the beamspace domain as
\begin{equation}
\bs{Q}_b = \bs{F} \bs{Q} \bs{F}^H,
\end{equation}
where $\bs{Q}_b$ represents the matrix $\bs{Q}$ in the beamspace domain. This transformation tends to concentrate the matrix energy into fewer significant coefficients, which results in the matrix $\bs{Q}_b$ being sparse. Such a sparse matrix can reduce the number of multiplications required for each
\ac{CG} iteration, which has the potential to decrease hardware
complexity and power dissipation, along with improved numerical efficiency~\cite{mirfarshbafan2020sparse}.


Preconditioned \ac{CG} (\ref{alg:pmatrix_cg_inverse}) is then applied to $\bs{Q}_b$ to compute the matrix $\bs{Q}_b^{-1}$. Finally, the result is transformed back to the antenna domain
\begin{equation}
\bs{Q}^{-1} \approx \bs{F}^H \bs{Q}_b^{-1} \bs{F}.
\end{equation}

  \subsection{Truncated Eigenvalue Decomposition} \label{sec:EVD}

We present a randomized \ac{EVD} method for complex-valued matrices, which provides a hardware-efficient truncated \ac{EVD} approach for estimating the $q$ largest eigenvalues and their corresponding eigenvectors of the matrix $\bs{Q}$. 
The proposed \ac{RC-EVD}, described in Algorithm~\ref{alg:randEVD}, is derived as a complex-valued extension of the randomized \ac{SVD} framework~\cite{martinsson2011randomized}. 
Furthermore, it is enhanced through subspace iterations (indexed by $p$) to improve approximation accuracy~\cite{tomas2023fast}. A rigorous error analysis of the \ac{RC-EVD} follows directly from the theoretical results in~\cite{halko2011finding}.


\begin{algorithm}[t]
\footnotesize
\caption{Randomized Complex EVD}
\label{alg:randEVD}
\begin{algorithmic}[1]

\Require Hermitian matrix $A \in \mathbb{C}^{N \times N}$, target rank $q \in [1,N]$, power iterations $p \geq 1$
\Ensure $U_q \in \mathbb{C}^{N \times q}, \ \Sigma_q = \mathrm{diag}(\sigma_1,\dots,\sigma_q)$

\State Generate a random matrix $Q_0 \in \mathbb{C}^{N \times q}$

\For{$j = 1, 2, \dots, p$}
    \State $Y_j \gets A Q_{j-1}$
    \State QRC: $Y_j = Q_j R_j$
\EndFor

\State $B \gets Q_p^H A Q_p$
\State Compute EVD: $B = \tilde{U} \Sigma_q \tilde{U}^H$
\State $U_q \gets Q_p \tilde{U}$

\end{algorithmic}
\end{algorithm}

There are two main non-trivial operations in the \ac{RC-EVD} algorithm. 
First, a small $q \times q$ eigenvalue decomposition is performed at the final stage. 
Since the complexity of this operation is $\mathcal{O}(q^3)$ and $q \ll N$, its computational cost is negligible compared to the overall algorithm. 
Furthermore, due to its small size and regular structure, this operation can be efficiently mapped onto hardware, particularly using systolic-array-based \ac{VLSI} architectures \cite{vishnu2026hardware}.

Second, the algorithm requires two QR factorizations involving “tall-and-skinny” matrices of dimensions $N \times q$. In the next section, we demonstrate how these QR factorizations can be efficiently implemented by a Cholesky-based orthogonalization scheme.
This reformulation significantly improves hardware efficiency by relying on structured, highly parallelizable operations that are well suited for systolic arrays.

\subsection{QRC: QR Factorization Using Cholesky Decomposition }
Conventional QR methods, such as Householder or Gram-Schmidt, involve irregular data access patterns and higher control complexity, making them less suitable for efficient hardware implementation~\cite{tomas2023fast}. Here, we employ QRC, a Cholesky-based QR decomposition method that leverages matrix multiplication and triangular solves to efficiently orthogonalize tall-and-skinny matrices using a small-scale Cholesky factorization~\cite{tomas2023fast}.
A complex-valued version of the QRC algorithm is introduced in Algorithm~\ref{alg:qrc}. As evident, there are two non trivial components in this algorithm:

\textbf{Solving a triangular linear system}:  
This operation corresponds to solving systems of the form $\bs{L}^H \bs{X} = \bs{B}$, where $\bs{L} \in \mathbb{C}^{q \times q}$ is triangular and $\bs{B} \in \mathbb{C}^{q \times N}$. 
From a hardware perspective, the operation exhibits a regular dataflow with a short sequential depth of $r$ steps and high parallelism across the $N$ columns, making it well suited for pipelined and systolic implementations~\cite{kung1979systolic}.

\textbf{Cholesky decomposition}: The core operation of the proposed QRC is a lightweight Cholesky decomposition, involving the factorization of a $q \times q$ matrix. This small size Cholesky decomposition is well suited for \ac{VLSI} implementation, particularly using systolic array architectures \cite{schreiber1986systolic, yin2013implementation}. 
 


\begin{algorithm}[t]
\footnotesize
\caption{QRC: QR factorization using Cholesky decomposition}
\label{alg:qrc}
\begin{algorithmic}[1]

\Require Matrix $Q \in \mathbb{C}^{q \times q}$
\Ensure Orthonormal $Q \in \mathbb{C}^{q \times q}$, upper triangular $R \in \mathbb{C}^{q \times q}$

\State $W \gets Q^H Q$
\State Compute Cholesky factorization: $W = L L^H$
\State $Q \gets Q L^{-H}$

\State $W \gets Q^H Q$
\State Compute Cholesky factorization: $W = \bar{L} \bar{L}^H$
\State $Q \gets Q \bar{L}^{-H}$

\State $R \gets L^H \bar{L}^H$

\end{algorithmic}
\end{algorithm}

\subsection{{Complexity Analysis}}
Table~\ref{tab:complexity} summarizes the computational cost of the proposed approaches. To further investigate the complexity, the following remark is noted:

\theoremstyle{remark}
\begin{remark} \label{rem:complexity}
The complexity of joint preconditioner construction and application in Table~\ref{tab:complexity} depends on the parameters $q$ and $p$ in Algorithm~\ref{alg:randEVD}. Based on \ac{VLSI}-oriented design considerations discussed throughout the paper, the parameter $q$ is restricted to small values (e.g., $q \in \{4,8\}$) to ensure the hardware scalability of \ac{RC-EVD} and Cholesky factorization. Similarly, the parameter $p$ is chosen as $p < 8$ to avoid excessive hardware overhead. A detailed simulation-based analysis of the impact of $q$ and $p$ is provided in the next section.

\end{remark}

 In general, the computational complexity of transforming to the beamspace domain is dominated by $\mathcal{O}(N^2 \log N)$, while the complexity of preconditioning is dominated by  $\mathcal{O}((p + k') N^2 q)$, depending on the choice of $p$ and $q$.
Consequently, as validated in simulations, both preconditioning and operation in the beamspace domain reduce the required number of \ac{CG} iterations from $k$ to $k' \ll k$. The computation saved by skipping $(k - k')$ iterations is $\mathcal{O}((k - k') N^3)$, which corresponds to the cost of standard \ac{CG}-based matrix operations. The additional overhead complexities of $\mathcal{O}(N^2 \log N)$ and $\mathcal{O}((p + k') N^2 q)$ are significantly lower than $\mathcal{O}((k - k') N^3)$. This demonstrates that the extra overhead introduced by beamspace transformation and preconditioning is negligible compared to the substantial computational savings achieved.



 
\begin{table}[ht]
\centering
\footnotesize
\caption{Computational cost of preconditioning and beamspace transformation in terms of complex-valued operations}
\label{tab:complexity}
\renewcommand{\arraystretch}{1.2}
\begin{tabular}{l||c|c}
\toprule
\textbf{Operation} & \textbf{Complexity} & \textbf{Remarks} \\
\midrule
\multicolumn{3}{c}{\textbf{QRC (per QR factorization)}} \\
\midrule
Multiplication $\bs{Q}^H \bs{Q}$ & $\mathcal{O}(N q^2)$ &  Dominant cost  \\
 Triangular  ($\bs{Q} \bs{L}^{-H}$) & $\mathcal{O}(N q^2)$ & Highly parallelizable \\
Cholesky factor.& $\mathcal{O}(q^3)$ & Negligible ($q \ll N$) \\
\hline
Dominated & $\mathcal{O}(N q^2)$ &  GEMM \\
\midrule
\multicolumn{3}{c}{\textbf{RC-EVD (Per iteration, $p \ll N$)}} \\
\midrule
Mul.s with form $\bs{A} \bs{Q}_{j}$ & $\mathcal{O}(N^2 q)$ & Dominant cost \\
QRC & $\mathcal{O}(N q^2)$ & Minor cost ($q \ll N$) \\
Small EVD & $\mathcal{O}(q^3)$ & Negligible \\
\hline
Dominated & $\mathcal{O}(N^2 q)$ & GEMM \\
\midrule
\multicolumn{3}{c}{\textbf{Joint Preconditioner Construction and Application}} \\
\midrule
RC-EVD & $\mathcal{O}(p N^2 q)$ & Dominant cost \\
Compute \ref{eq:impl} & $\mathcal{O}(N^2 q)$ & Per Iteration \\
\hline
Dominated ($k'$ Iter. ) & $\mathcal{O}((p+k') N^2 q)$ & see remark~\ref{rem:complexity}  \\
\midrule
\multicolumn{3}{c}{\textbf{Beamspace Transformation}} \\
\midrule
DFT ($\bs{F}\bs{Q}\bs{F}^H$) & $\mathcal{O}(N^2 \log N)$ & Done two times  \\
\bottomrule

\end{tabular}
\end{table}


 \section{Ray-Tracing Simulation Results} \label{results}

We evaluate the proposed framework on channels generated by the
NVIDIA Sionna ray tracer~\cite{hoydis2023sionna}, adopting the
propagation scenario and link parameters of~\cite{rasteh2025scalable}.
The per-user transmit powers are calibrated so that the post-beamforming
\ac{SNR} spans the range $[-6,\,14]$~dB, covering both noise-limited
and interference-limited operating points. The
base station is equipped with a $16 \times 16$ planar array; this
configuration is large enough to exhibit the eigenvalue clustering and
beamspace sparsity that drive the algorithmic gains reported in the
following subsections. The
extension to larger array geometries is left to the journal version of
this work.

\subsection{Performance Analysis of the Proposed Methods}

\begin{figure*}[!t]
\centering
\begin{subfigure}{0.25\textwidth}
    \centering
    \captionsetup{justification=centering}
    \includegraphics[width=\linewidth]
    {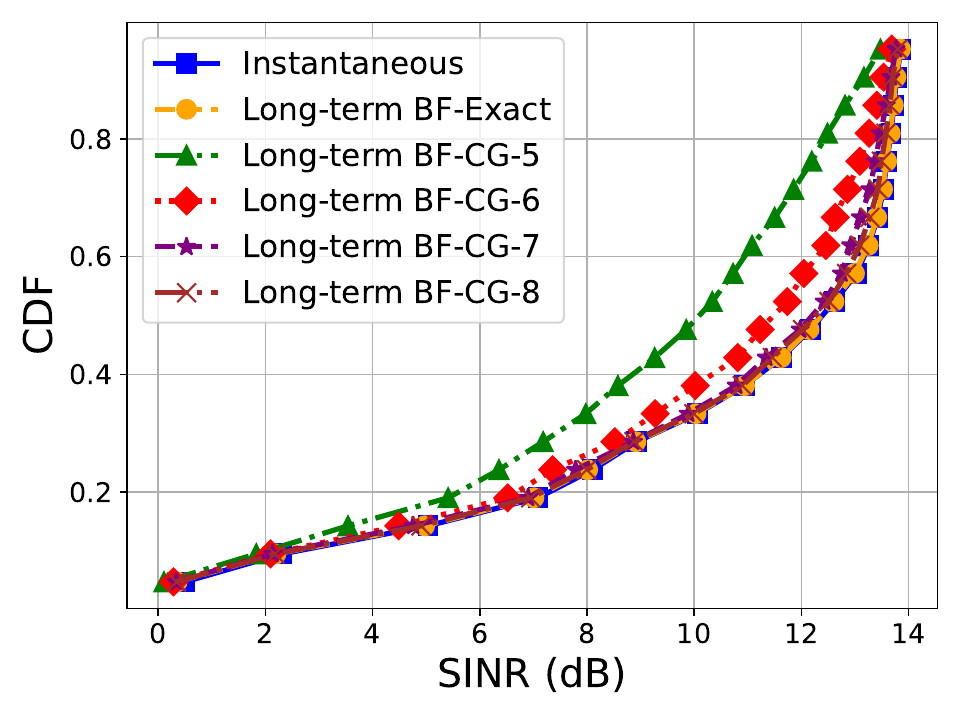}
    \caption{Beamspace domain\\With preconditioning}
        \label{fig:CDFPlotsa}

\end{subfigure}\hfill
\begin{subfigure}{0.25\textwidth}
    \centering
    \captionsetup{justification=centering}
    \includegraphics[width=\linewidth]{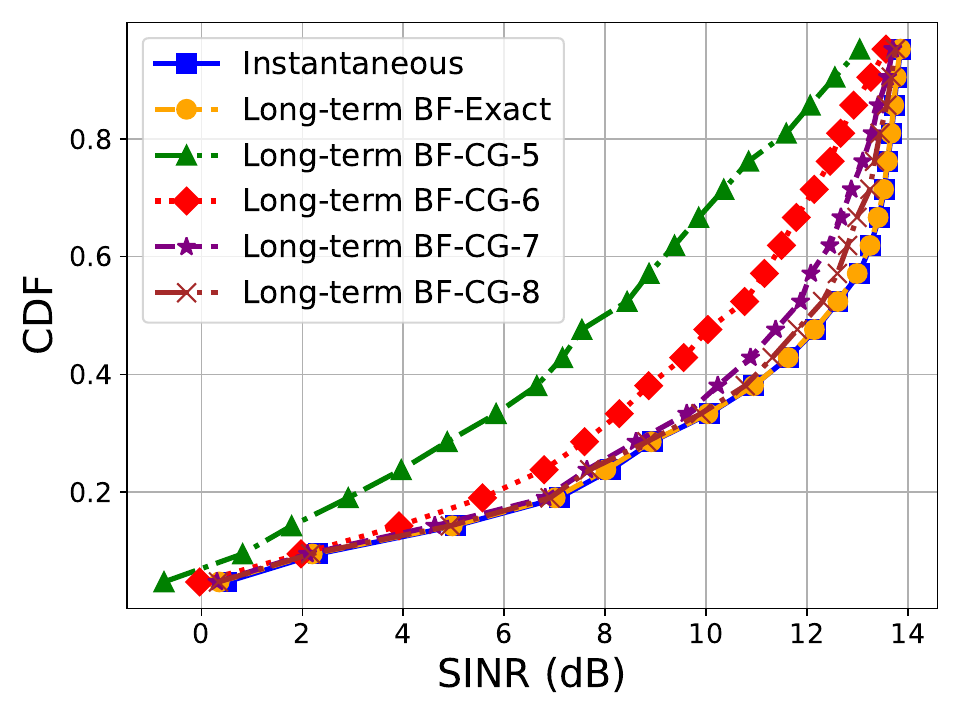}
    \caption{Antenna domain\\With preconditioning}
    \label{fig:CDFPlotsb}
\end{subfigure}\hfill
\begin{subfigure}{0.25\textwidth}
    \centering
    \captionsetup{justification=centering}
    \includegraphics[width=\linewidth]{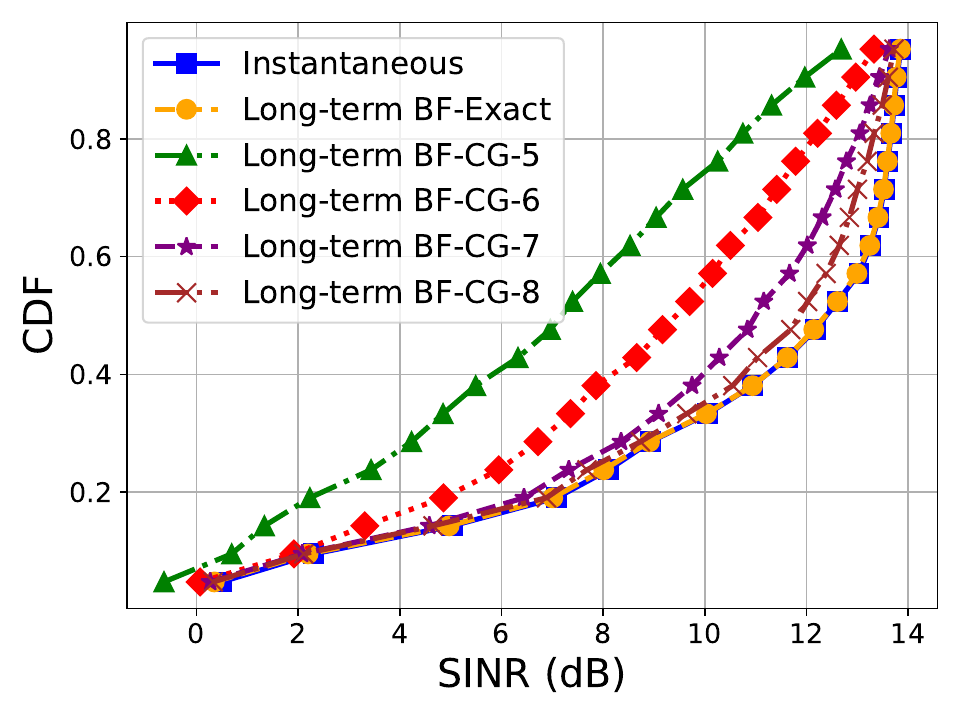}
    \caption{Beamspace domain\\Without preconditioning }
    \label{fig:CDFPlotsc}
\end{subfigure}\hfill
\begin{subfigure}{0.25\textwidth}
    \centering
    \captionsetup{justification=centering}
    \includegraphics[width=\linewidth]{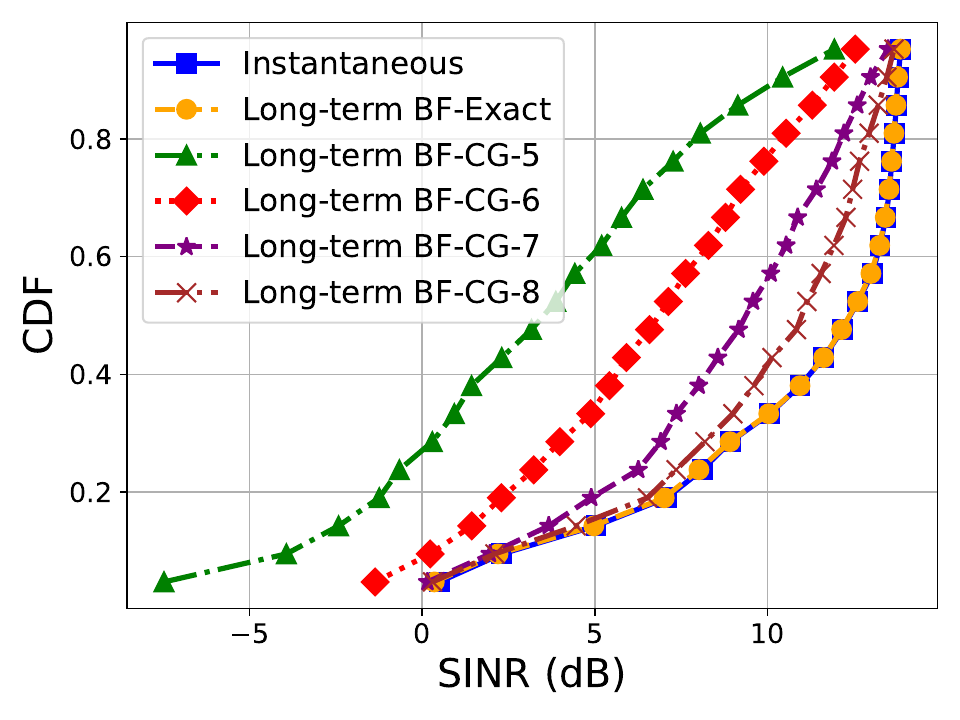}
    \caption{Antenna domain\\without preconditioning }
    \label{fig:CDFPlotsd}
\end{subfigure}
\caption{CDF of post-beamforming \ac{SINR}, showing the effect of operation in beamspace domain and employing preconditioning.}
\label{fig:CDFPlots}
\end{figure*}

Fig.~\ref{fig:CDFPlots} shows the \ac{CDF} of post-beamforming \ac{SINR} in different configurations. Note that in the figures, the optimal \ac{MMSE} beamformer \cite{rasteh2025scalable} and exact \ac{LTBF} (\ac{LTBF} in which matrix inversion is computed without any approximation) are included as baselines. 
A comparison of Figs.~\ref{fig:CDFPlotsd}--\ref{fig:CDFPlotsc} indicates that both preconditioning and operation in the beamspace domain shift the \ac{CDF} curves to the right, corresponding to improved performance. This improvement is more pronounced in the case of preconditioning. Furthermore, Fig.~\ref{fig:CDFPlotsd} presents the results obtained when preconditioning is applied in the beamspace domain, thereby combining the advantages of both approaches. This joint strategy yields a substantial enhancement in the performance of the matrix inversion approximation.

\begin{figure}
    \centering
    \includegraphics[width=0.95\linewidth]{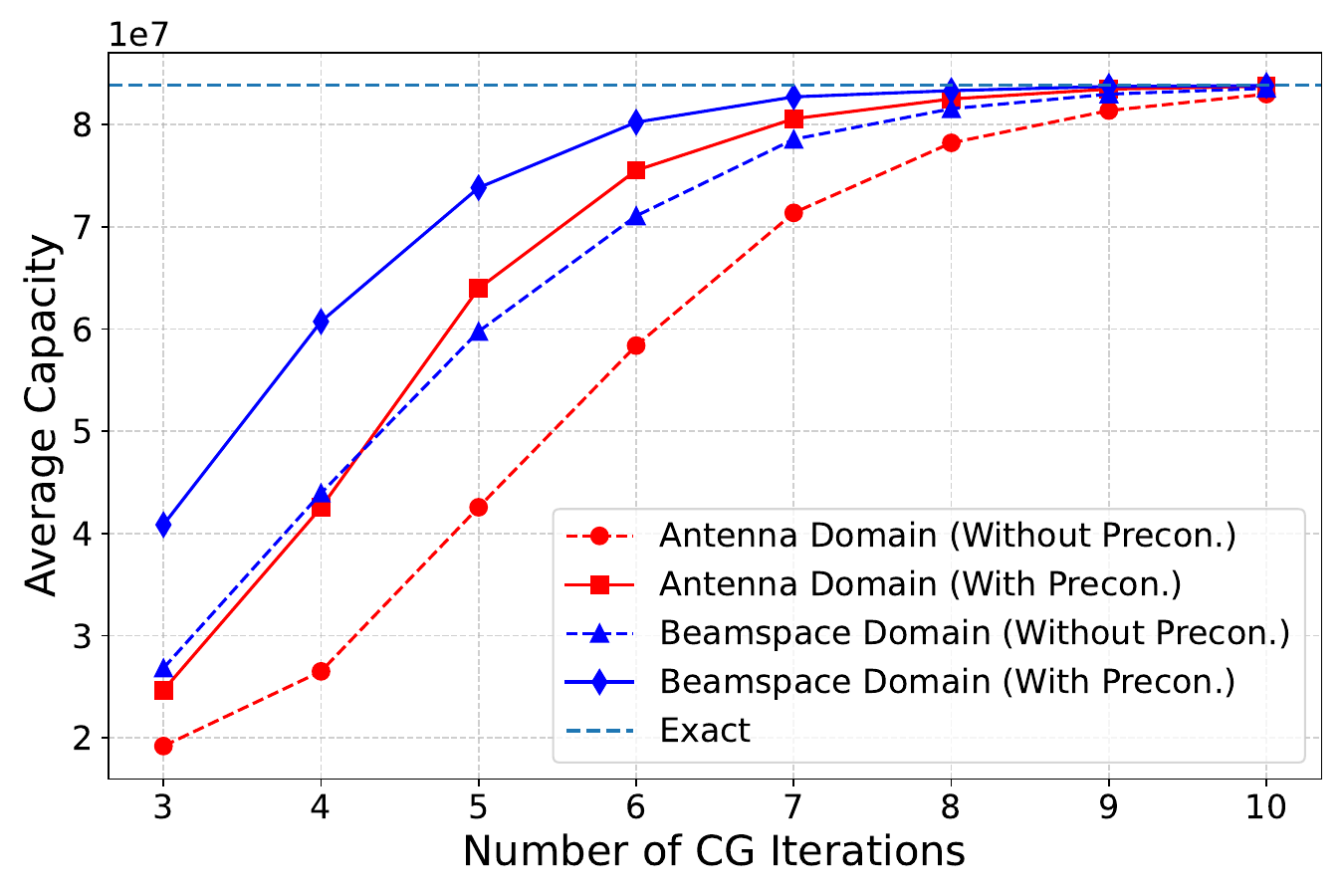}
    \caption{Average capacity versus number of CG iterations.
}
    \label{fig:cap}
\end{figure}

Fig.~\ref{fig:cap} illustrates the average capacity of \acp{UE} as a function of the number of \ac{CG} iterations. Overall, both preconditioning and operation in the beamspace domain reduce the required number of \ac{CG} iterations by approximately one. 
When the two approaches are combined, the capacity approaches the exact matrix inversion much faster than other methods. In particular, the required number of iterations is further reduced by about 2–3 iterations. Notably, in the low-iteration regime of the \ac{CG} algorithm (specifically at 6 and 7 iterations, where the combined method demonstrates satisfactory performance), the joint preconditioning and beamspace approach increases the capacity by more than 33\% and 16\%, respectively.  

\subsection{Sparsity of the Target Matrix}

To provide a qualitative overview of the impact of operating in the beamspace domain, a threshold of 0.005 is adopted solely for reporting the sparsity level of the matrices. It is important to note that the proposed method does not rely on, nor incorporate, any form of thresholding. Empirically, operation in the beamspace domain yields sparsity ratios of up to 55\%, while the corresponding matrices in the antenna domain exhibit sparsity levels below 2\%. 
These observations highlight the inherent sparsifying effect of the beamspace transformation. 

While the incorporation of explicit thresholding schemes could further enhance sparsity and thereby improve hardware efficiency, such approaches must be applied with caution. In particular, excessive sparsification may degrade the performance of the long-term beamformer. Moreover, improper thresholding can alter the numerical properties of the system matrix, potentially leading to instability or even divergence of the \ac{CG} algorithm.

\subsection{Complexity Trade-off in Choosing Parameters \texorpdfstring{$q$}{q} and \texorpdfstring{$p$}{p}}
\begin{figure}
    \centering
    \includegraphics[width=0.95\linewidth]{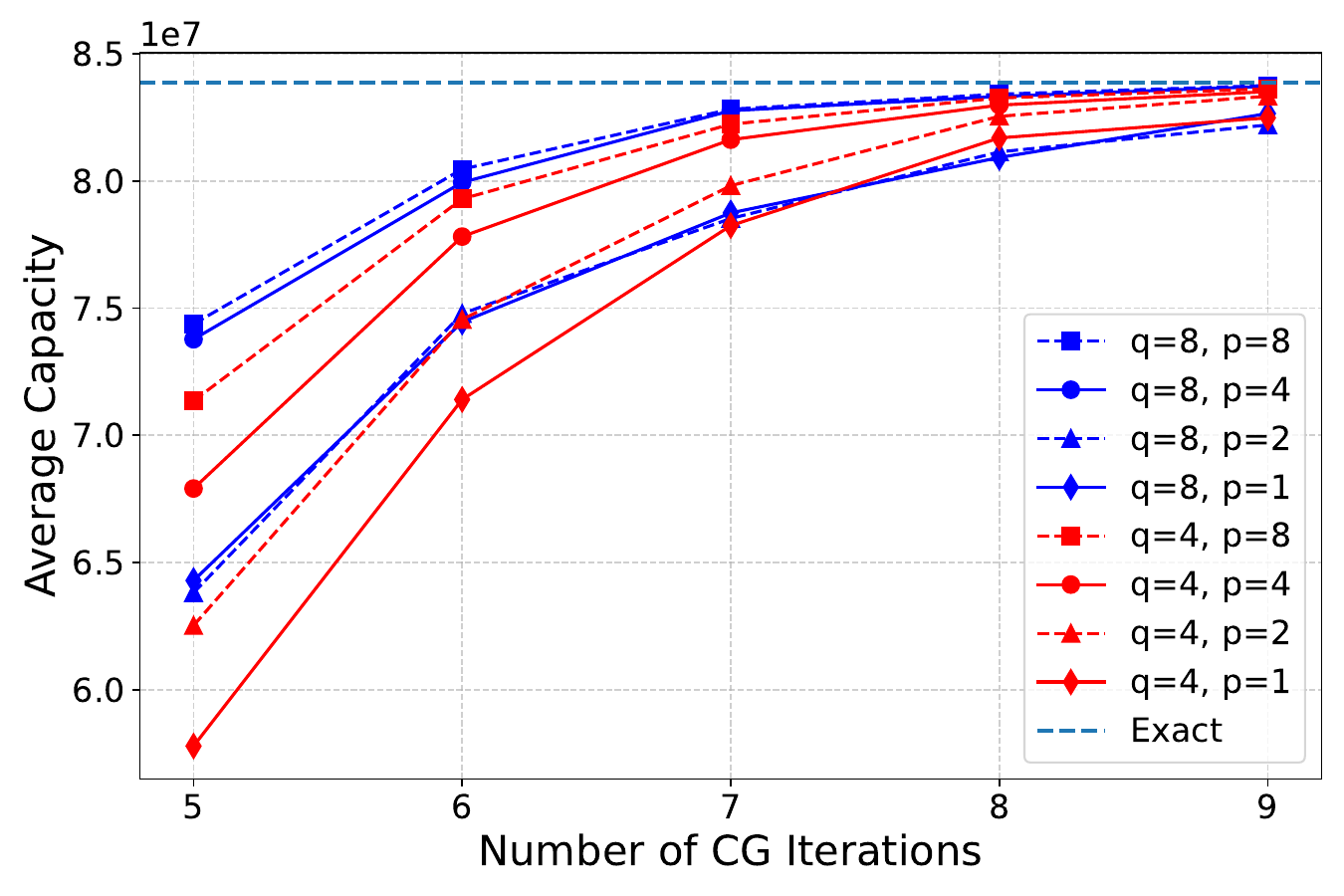}
    \caption{Average capacity versus number of CG iterations for different values of $p$ and $q$.
}
    \label{fig:rp}
\end{figure}

Fig.~\ref{fig:rp} illustrates the impact of different values of $p$ for $q = 4$ (red curves) and $q = 8$ (blue curves). As expected, increasing $q$ and $p$ leads to faster convergence toward the capacity achieved by the exact inversion of the matrix. However, the performance gap between $p = 4$ and $p = 8$ is marginal for the case $q = 8$, indicating diminishing returns for values of $p$ larger than 4. 
Another important observation is the necessity of selecting $p > 2$, as smaller values of $p$ result in a noticeable degradation in performance for both $q = 4$ and $q = 8$.
Based on this tradeoff, the results in Fig.~\ref{fig:CDFPlots} and~\ref{fig:cap} are generated using $q = 8$ and $p = 4$, which provide a favorable balance between performance and hardware efficiency.

 \section{Conclusion}
We have presented a hardware-friendly low-rank preconditioning
framework for the matrix inversion that underlies long-term
beamforming in massive \ac{MU-MIMO} systems. The preconditioner is
constructed at low cost via a randomized complex eigenvalue
decomposition, whose tall-and-skinny QR steps are realized through a
Cholesky-based factorization. These operations use matrix multiplications and triangular solves suited for systolic array architectures. Performing the
preconditioned \ac{CG} inversion in the beamspace domain provides an
additional, complementary convergence improvement, induced by the
inherent sparsification of the channel correlation in beamspace.
Ray-tracing simulations show that the proposed scheme reduces the \ac{CG}
iteration count by two to three relative to the unpreconditioned
baseline while preserving the post-equalization \ac{SINR} of the exact
inversion across the full operating range; at the low-iteration
regime most relevant for hardware deployment, the joint scheme
delivers up to $33\%$ capacity improvement over antenna-domain \ac{CG}
without preconditioning and $16\%$ over antenna-domain \ac{CG} with
preconditioning. Ongoing work addresses the extension to larger array
configurations, the integration of explicit sparsity-exploiting
arithmetic in the beamspace \ac{CG} kernel, and a full systolic-array
synthesis to translate the algorithmic complexity savings reported
here into measured energy and area gains on a baseband \ac{ASIC} target.


\bibliographystyle{refs/IEEEtran}
\bibliography{refs/IEEEfull,refs/bibliography}{}

@book{heath2018foundations,
  title={{Foundations of MIMO communication}},
  author={Heath Jr, Robert W and Lozano, Angel},
  year={2018},
  publisher={Cambridge University Press}
}

@inproceedings{lozano2007long,
  title={Long-term transmit beamforming for wireless multicasting},
  author={Lozano, Angel},
  booktitle={2007 IEEE International Conference on Acoustics, Speech and Signal Processing-ICASSP'07},
  volume={3},
  pages={III--417},
  year={2007},
  organization={IEEE}
}

@inproceedings{dai2021scalable,
  title={A scalable generator for massive MIMO baseband processing systems with beamspace channel estimation},
  author={Dai, Yue and Liew, Harrison and Rasekh, Maryam Eslami and Mirfarshbafan, Seyed Hadi and Gallyas-Sanhueza, Alexandra and Dunn, James and Madhow, Upamanyu and Studer, Christoph and Nikoli{\'c}, Borivoje},
  booktitle={2021 IEEE Workshop on Signal Processing Systems (SiPS)},
  pages={182--187},
  year={2021},
  organization={IEEE}
}

@inproceedings{sayeed2013beamspace,
  title={Beamspace MIMO for high-dimensional multiuser communication at millimeter-wave frequencies},
  author={Sayeed, Akbar and Brady, John},
  booktitle={2013 IEEE global communications conference (GLOBECOM)},
  pages={3679--3684},
  year={2013},
  organization={IEEE}
}

@techreport{nokia2025massiveMIMO,
  title        = {{Extreme Massive MIMO for Macro Cell Capacity Boost in 5G-Advanced and 6G}},
  author       = {Harri Holma, Harish Viswanathan and Preben Mogensen},
  year         = {2025},
  institution  = {Nokia},
  type         = {White Paper},
  url          = {https://www.nokia.com/asset/210786/}
}

@article{jin2023massive,
  title={{Massive MIMO evolution toward 3GPP release 18}},
  author={Jin, Huangping and Liu, Kunpeng and Zhang, Min and Zhang, Leiming and Lee, Gilwon and Farag, Emad N and Zhu, Dalin and Onggosanusi, Eko and Shafi, Mansoor and Tataria, Harsh},
  journal={IEEE Journal on Selected Areas in Communications},
  volume={41},
  number={6},
  pages={1635--1654},
  year={2023},
  publisher={IEEE}
}

@article{larsson2014massive,
  title={Massive MIMO for next generation wireless systems},
  author={Larsson, Erik G and Edfors, Ove and Tufvesson, Fredrik and Marzetta, Thomas L},
  journal={IEEE communications magazine},
  volume={52},
  number={2},
  pages={186--195},
  year={2014},
  publisher={IEEE}
}

@article{jia2025joint,
  title={{Joint Detection, Channel Estimation and Interference Nulling for Terrestrial-Satellite Downlink Co-Existence in the Upper Mid-Band}},
  author={Jia, Shizhen and Ying, Mingjun and Mezzavilla, Marco and Calin, Doru and Rappaport, Theodore S and Rangan, Sundeep},
  journal={arXiv preprint arXiv:2510.08824},
  year={2025}
}

@inproceedings{akrout2023bandwidth,
  title={{Bandwidth Gain: The Missing Gain of Massive MIMO}},
  author={Akrout, Mohamed and Shyianov, Volodymyr and Bellili, Faouzi and Mezghani, Amine and Heath, Robert W},
  booktitle={ICC 2023-IEEE International Conference on Communications},
  pages={5997--6003},
  year={2023},
  organization={IEEE}
}

@inproceedings{yin2015vlsi,
  title={VLSI design of large-scale soft-output MIMO detection using conjugate gradients},
  author={Yin, Bei and Wu, Michael and Cavallaro, Joseph R and Studer, Christoph},
  booktitle={2015 IEEE International Symposium on Circuits and Systems (ISCAS)},
  pages={1498--1501},
  year={2015},
  organization={IEEE}
}

@article{rasteh2025spatial,
  title={A Spatial Array for Spectrally Agile Wireless Processing},
  author={Rasteh, Ali and Hennessee, Andrew and Shivhare, Ishaan and Garg, Siddharth and Rangan, Sundeep and Reagen, Brandon},
  journal={arXiv preprint arXiv:2512.04182},
  year={2025}
}

@article{rasteh2025scalable,
  title={Scalable Long-Term Beamforming for Massive Multi-User MIMO},
  author={Rasteh, Ali and Kiani, Amirreza and Mezzavilla, Marco and Rangan, Sundeep},
  journal={arXiv preprint arXiv:2511.09464},
  year={2025}
}

@article{fang2025finite,
  title={Finite-Precision Conjugate Gradient Method for Massive MIMO Detection},
  author={Fang, Yiming and Chen, Li and You, Changsheng and Wen, Dingzhu and Zhu, Pengcheng},
  journal={arXiv preprint arXiv:2504.09820},
  year={2025}
}

@article{liu2020energy,
  title={Energy-and area-efficient recursive-conjugate-gradient-based MMSE detector for massive MIMO systems},
  author={Liu, Leibo and Peng, Guiqiang and Wang, Pan and Zhou, Sheng and Wei, Qiushi and Yin, Shouyi and Wei, Shaojun},
  journal={IEEE Transactions on Signal Processing},
  volume={68},
  pages={573--588},
  year={2020},
  publisher={IEEE}
}

@book{saad2003iterative,
  title={Iterative methods for sparse linear systems},
  author={Saad, Yousef},
  year={2003},
  publisher={SIAM}
}

@article{albreem2021low,
  title={Low complexity linear detectors for massive MIMO: A comparative study},
  author={Albreem, Mahmoud A and Salah, Wael and Kumar, Arun and Alsharif, Mohammed H and Rambe, Ali Hanafiah and Jusoh, Muzammil and Uwaechia, Anthony Ngozichukwuka},
  journal={IEEE Access},
  volume={9},
  pages={45740--45753},
  year={2021},
  publisher={IEEE}
}

@inproceedings{hoydis2023sionna,
  title={Sionna RT: Differentiable ray tracing for radio propagation modeling},
  author={Hoydis, Jakob and A{\"\i}t Aoudia, Fay{\c{c}}al and Cammerer, Sebastian and Nimier-David, Merlin and Binder, Nikolaus and Marcus, Guillermo and Keller, Alexander},
  booktitle={2023 IEEE Globecom Workshops (GC Wkshps)},
  pages={317--321},
  year={2023},
  organization={IEEE}
}

@article{vishnu2026hardware,
  title={A Hardware-Efficient QR Algorithm and Its VLSI Architecture for Eigenvalue Decomposition of Symmetric Matrices},
  author={Vishnu, PS and Francis, Jobin and Mula, Subrahmanyam},
  journal={IEEE Transactions on Very Large Scale Integration (VLSI) Systems},
  year={2026},
  publisher={IEEE}
}

@article{schreiber1986systolic,
  title={On systolic arrays for updating the Cholesky factorization},
  author={Schreiber, Robert and Tang, Wei-Pai},
  journal={BIT Numerical Mathematics},
  volume={26},
  number={4},
  pages={451--466},
  year={1986},
  publisher={Springer}
}

@inproceedings{yin2013implementation,
  title={Implementation trade-offs for linear detection in large-scale MIMO systems},
  author={Yin, Bei and Wu, Michael and Studer, Christoph and Cavallaro, Joseph R and Dick, Chris},
  booktitle={2013 IEEE international conference on acoustics, speech and signal processing},
  pages={2679--2683},
  year={2013},
  organization={IEEE}
}

@inproceedings{kung1979systolic,
  title={Systolic arrays (for VLSI)},
  author={Kung, Hsiang Tsung and Leiserson, Charles E and others},
  booktitle={Sparse Matrix Proceedings 1978},
  volume={1},
  pages={256--282},
  year={1979},
  organization={SIAM Philadelphia, PA, USA}
}

@article{martinsson2011randomized,
  title={A randomized algorithm for the decomposition of matrices},
  author={Martinsson, Per-Gunnar and Rokhlin, Vladimir and Tygert, Mark},
  journal={Applied and Computational Harmonic Analysis},
  volume={30},
  number={1},
  pages={47--68},
  year={2011},
  publisher={Elsevier}
}

@article{halko2011finding,
  title={Finding structure with randomness: Probabilistic algorithms for constructing approximate matrix decompositions},
  author={Halko, Nathan and Martinsson, Per-Gunnar and Tropp, Joel A},
  journal={SIAM review},
  volume={53},
  number={2},
  pages={217--288},
  year={2011},
  publisher={SIAM}
}

@article{tomas2023fast,
  title={Fast truncated SVD of sparse and dense matrices on graphics processors},
  author={Tom{\'a}s, Andr{\'e}s E and Quintana-Orti, Enrique S and Anzt, Hartwig},
  journal={The International Journal of High Performance Computing Applications},
  volume={37},
  number={3-4},
  pages={380--393},
  year={2023},
  publisher={SAGE Publications Sage UK: London, England}
}

@inproceedings{mirfarshbafan2020sparse,
  title={Sparse beamspace equalization for massive MU-MIMO mmWave systems},
  author={Mirfarshbafan, Seyed Hadi and Studer, Christoph},
  booktitle={ICASSP 2020-2020 IEEE International Conference on Acoustics, Speech and Signal Processing (ICASSP)},
  pages={1773--1777},
  year={2020},
  organization={IEEE}
}

@article{diouane2026spectral,
  title={A Spectral Preconditioner for the Conjugate Gradient Method with Iteration Budget},
  author={Diouane, Youssef and G{\"u}rol, Selime and Mouhtal, Oussama and Orban, Dominique},
  journal={arXiv preprint arXiv:2603.28969},
  year={2026}
}


\end{document}